%% file: main.tex
\documentclass{Interspeech}

\interspeechcameraready

\title {Lightweight Front-end Enhancement for Robust ASR via Frame Resampling and Sub-Band Pruning
\thanks{$^{\dagger}$Yanmin Qian is corresponding author}
}

\author[affiliation={1}]{Siyi}{Zhao}
\author[affiliation={1}]{Wei}{Wang}
\author[affiliation={1,2}]{Yanmin}{Qian$^{\dagger}$}

\affiliation{}{}{$^1$Auditory Cognition and Computational Acoustics Lab}
\affiliation{}{}{MoE Key Lab of Artificial Intelligence, AI Institute}
\affiliation{}{}{School of Computer Science, Shanghai Jiao Tong University, Shanghai,China}
\affiliation{}{}{$^2$Suzhou Institute of Artificial Intelligence, Shanghai Jiao Tong University, Suzhou,China}
\email{\{zsy\_coding,wangwei.sjtu,yanminqian\}@sjtu.edu.cn}
\keywords{speech enhancement, robust speech recognition, frame resampling, sub-band pruning}

\usepackage{comment}
\usepackage{subfiles}
\usepackage{url}
\usepackage{adjustbox}
\usepackage{algorithmic}
\usepackage{algorithm}

\begin{document}

\maketitle

\begin{abstract}
    
    Recent advancements in automatic speech recognition (ASR) have achieved notable progress, whereas robustness in noisy environments remains challenging. While speech enhancement (SE) front-ends are widely used to mitigate noise as a preprocessing step for ASR, they often introduce computational non-negligible overhead. This paper proposes optimizations to reduce SE computational costs without compromising ASR performance. Our approach integrates layer-wise frame resampling and progressive sub-band pruning. Frame resampling downsamples inputs within layers, utilizing residual connections to mitigate information loss. Simultaneously, sub-band pruning progressively excludes less informative frequency bands, further reducing computational demands. Extensive experiments on synthetic and real-world noisy datasets demonstrate that our system reduces SE computational overhead over 66\% compared to the standard BSRNN, while maintaining strong ASR performance.
\end{abstract}

\section{Introduction}
\label{sec:intro}

\subfile{sections/intro}

\begin{figure*}[!tp]
  \centering
  \includegraphics[width=\linewidth]{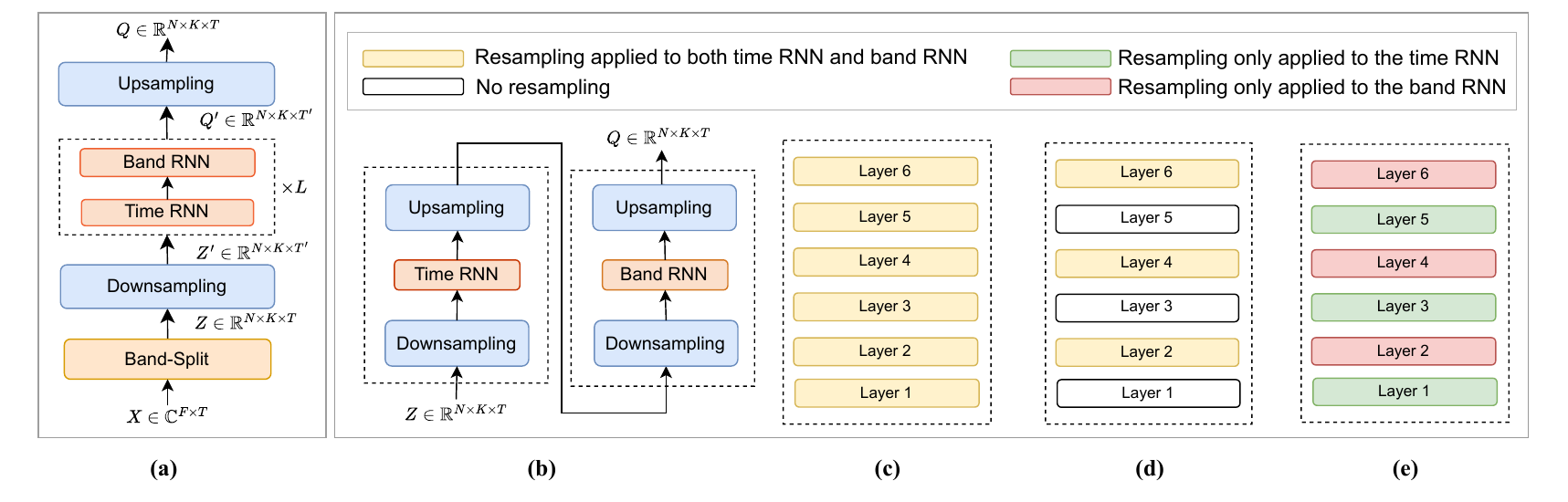}
  \caption{Proposed resampling methods: (a) Pre-downsampling and post-upsampling. (b) A single layer with resampling applied to both time and band RNNs. (c)-(e) Layer-wise resampling strategies as outlined in Section \ref{where}. Resampling for both the time RNN and the band RNN is applied along dimension (\(T\)).}.
  \label{fig:subsample}
\end{figure*}

\section{Methodology}

\subsection{Revisit on Band-Split RNN~(BSRNN)}
\label{sec:methods:bsrnn}
\subfile{sections/methods/bsrnn.tex}

\subsection{Frame and Sub-Band Resampling}
\label{sec:methods:subsample}
\subfile{sections/methods/subsample}

\subsection{Sub-band Pruning}
\label{sec:methods:skip}
\subfile{sections/methods/skip_v2}

\section{Experiment Setup}
\label{sec:exp_setup}
\subfile{sections/experiment_setup}

\section{Results and Analysis}
\label{sec:ret}
\subfile{sections/results}

\section{Conclusion}
In this paper, we proposed a set of optimizations to reduce computational cost of SE front-ends without compromising ASR performance. Our methods involve layer-wise frame resampling and sub-band pruning. We investigated different frame resampling strategies and leverage residual connections to mitigate potential information loss. Additionally, sub-band pruning progressively skips computations for less informative sub-bands, further enhancing computational efficiency. Grouped RNNs were employed to further reduce computational complexity without impacting performance. Through extensive experiments on synthetic and real-world datasets, we demonstrated that the system with these optimizations reduces the computational load over 66\% compared to the standard BSRNN, while limits the performance decline to relative 1\% across different ASR back-ends and datasets.

\newpage

\section{Acknowledgement}
This work was supported in part by the Key Research and Development Program of Jiangsu Province, China (Grant No. BE2022059-4), in part by China NSFC projects under Grants 62122050 and 62071288, and in part by Shanghai Municipal Science and Technology Commission Project under Grant 2021SHZDZX0102.

\bibliographystyle{IEEEtran}
\bibliography{mybib}

\end{document}

%% file: sections/intro.tex
Speech enhancement (SE) is essential for improving speech intelligibility and quality in acoustically challenging environments~\cite{f1, f2, f3, t1, t2, t3, t4}. As a preprocessing front-end, SE is widely used to enhance the noise robustness of automatic speech recognition (ASR) systems, a necessity for real-world applications~\cite{radford2023robust, hu2022interactive, ww, zhang2018deep, liu2019jointly, wang2022text, wang2022wav2vec}. While deep learning has driven significant improvements in SE performance over traditional methods, these models often rely on large architectures and increased complexity to achieve better results. This, however, leads to substantial computational demands, which hinders the practical deployment of SE models in resource-constrained ASR systems.

Reducing computational complexity for real-time applications is a key focus in SE research~\cite{fedorov20_interspeech, tan2021towards, lee2021demucs, choi2021real, braun2021towards, li2022skim}. Techniques such as quantization and pruning have been employed to reduce model size and complexity in existing architectures~\cite{fedorov20_interspeech, tan2021towards}. Other approaches focus on designing more efficient models specifically for low-computation inference~\cite{choi2021real, braun2021towards, li2022skim}, while recent efforts~\cite{hao2022fast, rong2024gtcrn} modify modules within high-performance SE models to further improve efficiency. Although these methods have demonstrated substantial reductions in computational cost while preserving enhancement quality, their effectiveness as preprocessing front-ends for ASR systems has not been thoroughly evaluated. Preliminary findings indicate that these lightweight SE models often fail to improve ASR performance when deployed as preprocessing front-ends, with this challenge commonly attributed to the speed-distortion phenomenon~\cite{oa, ochiai2024rethinking}.
In this work, we propose a lightweight SE model based on band-split RNN (BSRNN)~\cite{bsrnn} as a denoising frontend for ASR systems constrained by resource limitations. BSRNN was selected as the backbone due to its demonstrated ability to maintain high performance under constrained computational resources~\cite{beyond}. Our preliminary experiments also show that BSRNN introduces minimal distortion to the enhanced signal and improves ASR accuracy. To further reduce computational complexity and enable efficient processing on edge devices, we propose to incorporate resampling and sub-band pruning, optimizing the model for deployment as a preprocessing front-end in ASR applications. For resampling, we explore a range of temporal and sub-band resampling strategies to reduce computational costs while minimizing information loss to avoid performance degradation. Regarding sub-band pruning, we propose progressively skipping the computation of higher frequency sub-bands, as critical speech information predominantly resides in the lower and mid-range frequencies, with higher frequencies contributing less to intelligibility~\cite{ansi1997methods, kryter1962methods}. To demonstrate compatibility with other efficiency-enhancing techniques, we integrate grouped RNN~\cite{gao2018efficient} as a replacement for standard RNN layers, further reducing computational costs. To optimize the enhanced signal for ASR, we apply the observation adding (OA)~\cite{oa} technique, which involves adding a scaled version of the original signal to the enhanced output to mitigate distortions and artifacts. By incorporating these techniques, we achieve a substantial reduction in computational cost without compromising ASR performance.
Our contributions can be summarized as follows:
\begin{enumerate}
    \item We propose a layer-wise resampling strategy implemented in a staggered fashion, effectively reducing computational cost while minimizing information loss.
    \item We introduce a progressive sub-band pruning method to further decrease computational requirements by selectively skipping higher frequency sub-bands.
    \item Experiments on synthetic and real-world datasets show that our approach reduces SE model computational cost over 66\% while preserving robust ASR performance.
\end{enumerate}

%% file: sections/methods/bsrnn.tex
We begin by briefly introducing BSRNN\cite{bsrnn}, which serves as the backbone of our method. It is a frequency-domain model that splits the input spectrogram into subbands with predefined bandwidths and uses stacked RNN layers for cross-band and sequence modeling.


Given the input mixture, BSRNN first separates the complex-valued spectrogram \(X\) into \(K\) sub-bands with predefined bandwidths. For each sub-band spectrogram \(B_i\), a real-valued feature \(Z_i\) is created by concatenating its real and imaginary parts. These sub-band features are then merged into a full-band feature \(Z\). BSRNN performs sequence and cross-band modeling along the time dimension \(T\) and band dimension \(K\) through \(L\) iterations, producing an output \( Q \in \mathbb{R}^{N \times K \times T} \). Each of the \(K\) bands is then processed through band-specific fully connected (FC) layers, and the resulting features are concatenated to generate the predicted complex-valued time-frequency mask \(M \in \mathbb{C}^{F \times T}\) and the residual spectrogram \(R \in \mathbb{C}^{F \times T}\). The final target spectrogram \(S\) is computed by applying the Hadamard product of \(M\) and \(X\), then adding the residual spectrogram \(R\).
\begin{equation}
\bar{S} = M \odot X + R
\end{equation}

%% file: sections/methods/subsample.tex

Resampling, including both downsampling and upsampling, is a widely used technique for reducing the computational complexity of SE models. In this section, we investigate various resampling strategies, as illustrated in Fig~\ref{fig:subsample}, aiming to lower computational demands while minimizing the loss of crucial information.


\subsubsection{Pre-Downsampling and Post-Upsampling~(PPS)}
As shown in Fig~\ref{fig:subsample}(a), a straightforward approach is to apply downsampling before the input enters the network and upsampling after the network processes the data. While this reduces the number of frames and sub-bands that the network must handle, thus lowering computational complexity, it incurs considerable information loss. A substantial portion of the input is never processed by the network, leading to the omission of important information.

\subsubsection{Layer-Wise Resampling~(LWR)}
\label{where}
LWR selectively applies resampling to specific layers, ensuring that the output feature shape remains consistent with the input for each chosen layer. Unlike PPS, which affects the entire network with information loss, LWR confines the impact to selected layers. By utilizing residual connections, other layers still retain access to the full, intact input. This approach strikes a balance between computational efficiency and the preservation of critical information. We propose three LWR strategies, outlined as follows:

\begin{enumerate}
    \item LWR-ALL: As illustrated in Fig~\ref{fig:subsample}(c), this strategy applies downsampling before and upsampling after each group of time and band RNNs, ensuring a consistent feature shape while reducing computational load across both dimensions.
    \item LWR-SYNC: As shown in Fig~\ref{fig:subsample}(d), LWR-SYNC applies resampling to both time and band RNNs, but only in the odd-numbered layers, leaving the even-numbered layers unaltered to help retain more of the original input information.
    \item LWR-ASYNC: As depicted in Fig~\ref{fig:subsample}(e), LWR-ASYNC alternates resampling between time RNNs and band RNNs, applying it to time RNNs in odd layers and band RNNs in even layers. This approach distributes the potential information loss more evenly throughout the network.
\end{enumerate}

Resampling for both time and band RNNs follows the same approach, differing only in the dimension along which it is applied. Taking resampling for the time RNN as an example, the input frames are divided into blocks of size \(S\) along the temporal dimension \(T\). The mean value of each block is computed, resulting in an output \(Z' \in \mathbb{R}^{N \times K \times T'}\), where \(T' = \lceil T/S \rceil \). For upsampling, each feature in \(Z'\) is repeated \(S\) times along the temporal dimension, restoring the original time resolution. The same process is applied for the band RNN.




%% file: sections/methods/skip_v2.tex
\begin{figure}[htb!]
  \centering
  \includegraphics[width=\linewidth]{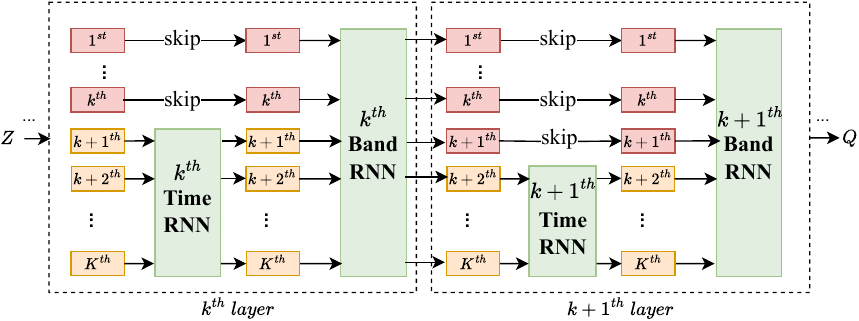}
  \caption{Illustration of the progressive subband pruning (SBP-P) in grouped BSRNN: All subbands are processed through Band RNNs and sorted from high to low frequency, with the number of subbands passing through Time RNNs progressively decreasing. Yellow blocks represent subbands processed by Time RNNs, while red blocks indicate skipped subbands.}
  \label{fig:skip}
\end{figure}

We propose a sub-band pruning (SBP) strategy that selectively skips computation of the higher frequency bands. Since most critical speech information is concentrated in the lower and mid-range frequencies, while higher frequencies contribute less to intelligibility~\cite{ansi1997methods, kryter1962methods}, this approach reduces computational complexity with minimal performance degradation. By reallocating resources to the more relevant frequency bands, SBP improves efficiency without compromising speech enhancement quality. We present two SBP strategies as follows:

\begin{enumerate}
    \item Aggressive (SBP-A): This strategy skips computing the highest \(L\) sub-bands right from the first time RNN layer, reducing computational load early in the process.
    
    \item Progressive (SBP-P): In this strategy, one additional high-frequency sub-band is progressively skipped as each new time RNN layer is entered as illustrated in Fig~\ref{fig:skip}. 
\end{enumerate}

\subsection{Revisit on Grouped RNN with Rearrangement}
To further optimize computational efficiency, we replace the standard RNN layers with Grouped RNN (GR) layers~\cite{gao2018efficient}, employing a group size of 2. In this approach, each GR layer divides both the input sequences and hidden states into two disjoint groups. Recurrent learning is performed independently within each group to model intra-group dependencies. To capture inter-group correlations, we introduce a rearrangement layer between consecutive recurrent layers and time steps, enabling the exchange of information across groups. This design improves computational efficiency while maintaining the model’s ability to capture essential temporal dependencies.

%% file: sections/experiment_setup.tex
\subsection{Datasets}
We conduct experiments on both English and Mandarin datasets. For English experiments, an online simulation combining clean speech from the \texttt{tr\_05\_simu\_1\_ch} subset of the \textsc{CHiME-4} corpus~\cite{chime4} with noise samples from the DNS dataset~\cite{dns} is used to generate the training data, across SNR levels from -5 dB to 20 dB. For testing, the official validation and test sets from \textsc{CHiME-4}, specifically \texttt{dt\_05\_real} and \texttt{et\_05\_real} 1ch track, are used. As for \textsc{DNS} dataset, it includes recordings from over 10,000 speakers reading 10,000 books at a 16kHz sampling rate. We generated 100 hours of noisy speech following the ESPnet\cite{esp} recipe, using the official validation set and test sets with two categories of synthetic clips: with and without reverberation.

For the Mandarin experiments, we combine clean speech from the \texttt{train} and \texttt{dev} sets of the \textsc{Aishell-1} corpus with noise from the \textsc{DNS} dataset at SNR levels from -5 dB to 20 dB to form new training and development sets. The in-house recorded dataset employed for evaluation consists of two real-world acoustic environments: a kitchen and a shopping mall. Utilizing a mobile phone with dual microphones positioned at the top and bottom, recordings were captured in both environments, while the shopping mall environment exhibits considerably greater reverberation compared to the kitchen. Each setting includes three test sets, with each set containing 1.5 hours of data evenly distributed across the two settings.

\subsection{Training Configuration}
All experiments were conducted using the ESPnet toolkit \cite{esp}. For SE front-ends, we adhered to the band-splitting configuration \textit{V4} outlined in \cite{bsrnn}, producing a total of 23 sub-bands. Our SE front-ends were trained for 54K steps, utilizing the Adam optimizer \cite{adam} and criteria included multi-resolution STFT magnitude loss \cite{m1} and SI-SNR loss. For DNS and \textsc{CHiME-4} datasets, we utilized the Whisper models\footnote{\url{huggingface.co/openai}} as the ASR back-ends. Notably, the Whisper models utilized in the experiments were not fine-tuned on the CHIME-4 dataset. For experiments on Mandarin dataset, we employed an officially released Paraformer model\footnote{\url{huggingface.co/funasr/Paraformer-large}}\cite{para}, renowned for its strong performance in Mandarin ASR tasks.


\subsection{Decoding}
Artifact errors from single-channel SE front-ends may cause ASR performance decline in noisy environments, and thus observation adding (OA) technique\cite{oa} was employed. OA adds a scaled observed signal to the enhanced signal, formulated as \(
    S_{OA} = \omega_{OA} \cdot S_{noisy} + (1 - \omega_{OA}) \cdot S_{enh}\), where $\omega_{OA} \in [0,1]$.  The $\omega_{OA}$ values are tuned separately on validation set in all experiments.
During the decoding process, we did not incorporate any language models, but applied text normalization\footnote{\url{huggingface.co/openai/whisper-large-v2\#evaluation}} to the decoding results of Whisper models.

\subsection{Evaluation Metrics}
We evaluate the model performance using Word Error Rate (WER), a standard metric in speech recognition tasks. Additionally, we quantify the computational cost of the proposed model adaptations using the number of Multiply-Accumulate Operations (MACs). This metric provides a hardware-agnostic measure of model complexity, allowing for a clear comparison of the computational efficiency of different adaptations. To simplify the representation of numerical values, the computational cost is reported in MACs (G/s), where 1 G/s denotes $10^9$ MACs per second, equivalent to 1 GMAC/s. This notation is adopted consistently throughout the results and tables for enhanced clarity and conciseness.

%% file: sections/results.tex
\subsection{Comparison of Different Resampling Strategies}


The results in Table \ref{tab:dns} indicate that LWR-ALL and PPS have the same computational cost, with LWR-ALL(4) referring to the application of the LWR-ALL strategy with a resampling factor of 4. However, both methods show a significant performance degradation compared to the original SE model. This decline in performance is likely due to the overly aggressive resampling strategy, which leads to substantial information loss.

In contrast, LWR-SYNC and LWR-ASYNC outperform the previous methods by applying resampling only to selected target layers, as shown in Table \ref{tab:dns}, while maintaining performance comparable to the original SE front-end. Notably, LWR-ASYNC surpasses LWR-SYNC across all ASR back-ends on both the \texttt{Without Reverb} and \texttt{With Reverb} test sets. This suggests that interleaving resampling across the temporal and band dimensions more effectively preserves critical speech information, leading to preserved performance.

We examine the impact of the resampling factor \( S \) for LWR on computational cost. As shown in Table~\ref{tab:dns}, comparing configurations 2, 4d, and 4e, a resampling factor of 4 significantly reduces the computational cost, while increasing the factor to 16 yields only marginal additional savings. This is because the interleaved resampling strategy inherently limits the maximum computational cost reduction to 50\%. Moreover, comparisons between LWR-ASYNC(4) and LWR-ASYNC(16) across all tables indicate that a higher resampling factor of 16 does not negatively affect ASR performance. In contrast, using a synchronized resampling strategy, as in LWR-SYNC(4), does lead to performance degradation, highlighting the advantage of the LWR-ASYNC approach.

\setlength{\tabcolsep}{3pt}
\begin{table}[htb!]
    \centering
    \normalsize
    \caption{WER(\%) on \textsc{DNS} challege test dataset with three sizes of Whisper models.}
    \begin{adjustbox}{max width=\linewidth}
     \begin{tabular}{lcccccccc}
    \toprule
    Method & \#MACs(G/s) & \multicolumn{3}{c}{Without Reverb} & \multicolumn{3}{c}{With Reverb}\\
    \cmidrule(lr){3-5} \cmidrule(lr){6-8}
     &  & Tiny & Medium & Large & Tiny & Medium & Large & \\
    \midrule
    1. Noisy  & —— & 15.02 & 8.85 & 4.80 & 39.51 & 21.13 & 10.94 \\
    \midrule
    2. BSRNN  & 1.84 & 11.36 & 5.59 & 4.23 & 36.60 & 15.34 & 10.29 \\
    \midrule
    3. +GR  & 1.09 & 11.75 & 5.72 & 4.40 & 36.82 & 15.60 & 10.87 \\
    \midrule
    4a. + PPS & 0.55 & 13.63 & 5.84 & 4.95 & 38.19 & 20.93 & 10.91 \\ 
    4b. + LWR-ALL(4) & 0.55 & 13.82 & 5.92 & 4.67 & 38.84 & 19.42 & 10.91 \\ 
    4c. + LWR-SYNC(4) & 1.19 & 11.93 & 5.77 & 4.38 & 36.72 & 15.56 & 10.89 \\
    4d. + LWR-ASYNC(4) & 1.19 & 11.49 & 5.67 & 4.25 & 36.67 & 15.42 & 10.59 \\
    4e. + LWR-ASYNC(16) & 1.03 & 11.74 & 5.68 & 4.33 & 36.80 & 15.49 & 10.62 \\
    \midrule
    5a. ++SBP-A & 0.96 & 11.81 & 5.84 & 4.38 & 38.02 & 15.74 & 11.41 \\ 
    5b. ++SBP-P & 0.99 & 11.76 & 5.69 & 4.35    & 36.82 & 15.59 & 10.87 \\ 
    \midrule
    6. +++GR & 0.62 & 11.76 & 5.72 & 4.40 & 36.87 & 15.61 & 10.90 \\ 
    \bottomrule
    \end{tabular}
\end{adjustbox}
    \label{tab:dns}
    \vspace{-2mm}
\end{table}


\subsection{Performance of Sub-band Pruning}



We validate the proposed sub-band pruning strategy on both English and Mandarin test sets using Whisper models.

To further evaluate our proposed methods and their variants, we combined LWR-ASYNC (with a downsample size of 16) with both SBP-A and SBP-P strategies. As shown in Table \ref{tab:dns}, SBP-P effectively limits performance degradation to no more than a 1\% relative decrease across all test sets, a trend that is consistent in Tables \ref{tab:chime} and \ref{tab:honor}. These results demonstrate that SBP-P efficiently reduces computational overhead while maintaining performance comparable to the original SE model. In contrast to SBP-A, which employs a more aggressive pruning strategy, SBP-P adopts a progressive reduction of sub-bands across layers. This approach leads to improved ASR results compared to SBP-A, with only a minimal GMAC difference of 0.03.



\begin{table}[htb!]
\setlength{\tabcolsep}{3pt}
    \centering
    \normalsize
    \caption{WER(\%) on \textsc{CHiME-4} 1-channel real noisy test sets with three sizes of Whisper models.}
    \begin{adjustbox}{max width=\linewidth}
     \begin{tabular}{l|c|ccc|ccc}
    \specialrule{1.2pt}{0pt}{0pt}
    Method & \#MACs(G/s) & \multicolumn{3}{c|}{dev} & \multicolumn{3}{c}{eval}\\
    \cmidrule(lr){3-5} \cmidrule(lr){6-8}
     &  & Tiny & Medium & Large & Tiny & Medium & Large \\
    \midrule
    1. Noisy  & —— & 21.76 & 5.55 & 4.41 & 35.03 & 8.63 & 6.69  \\
    \midrule
    \midrule
    2. DPARN\cite{dparn}  & 1.22 & 21.13 & 5.52 & 4.67 & 32.97 & 8.84 & 7.56 \\
    \midrule
    \midrule
    3. BSRNN  & 1.84 & 16.50 & 5.19 & 4.25 & 26.24 & 8.17 & 6.36 \\
    \midrule
    4. +GR  & 1.09 & 16.70 & 5.30 & 4.35 & 27.07 & 8.28 & 6.40 \\
    \midrule
    5a. + LWR-ASYNC(4) & 1.19 & 16.68 & 5.22 & 4.28 & 26.64 & 8.18 & 6.37 \\
    5b. + LWR-ASYNC(16) & 1.03 & 16.68 & 5.26 & 4.34 & 26.66 & 8.24 & 6.41 \\
    \midrule
    6. ++SBP-P & 0.99 & 16.69 & 5.28 & 4.34 & 26.66 & 8.24 & 6.57 \\ 
    \midrule
    7. +++GR & 0.62 & 16.72 & 5.31 & 4.35 & 27.12 & 8.30 & 6.40\\
    \specialrule{1.2pt}{0pt}{0pt}
    \end{tabular}
\end{adjustbox}
    \label{tab:chime}
    \vspace{-2mm}
\end{table}


Furthermore, experiments with the standalone GR configuration yielded a GMAC value of 1.09, indicating a higher computational cost compared to the proposed method. Despite the increased overhead, the performance of the GR configuration does not match that of the proposed methods, which consistently outperform GR across all three datasets. We also tested the results of the open-source lightweight SE model DPARN\cite{dparn} on widely-used public datasets \textsc{CHiME-4}. Although the GMAC value of DPARN exceeds that of the proposed methods, its ASR performance is inferior. This discrepancy can be attributed to the misalignment between SE and ASR metrics, as they do not always correlate. This underscores the objective of our proposed methods, which is to reduce the computational cost of SE while evaluating their effectiveness in robust ASR scenarios. In practice, lightweight SE models that perform well in SE metrics often exhibit suboptimal ASR performance.

Notably, the proposed methods are compatible with GR, and the performance is nearly equivalent to that of the standalone GR configuration, while maintaining significantly lower computational cost. We combined LWR-ASYNC with SBP-P and GR, as shown in the last row of Tables \ref{tab:dns}, \ref{tab:chime}, and \ref{tab:honor}. The results demonstrate that the GR configuration achieves a GMAC value as low as 0.62, representing a substantial reduction compared to the original models. Despite this considerable reduction in computational overhead, performance remains robust, with only minor degradation observed on certain subsets of the in-house Mandarin test sets.


\setlength{\tabcolsep}{3pt} 
\begin{table}[htb!]
    \centering
    \normalsize 
    \caption{CER(\%) on in-house Mandarian test sets.}
    \begin{adjustbox}{max width=\linewidth}
     \begin{tabular}{l|c|cccc|ccccc}
    \specialrule{1.2pt}{0pt}{0pt}
    Method & \#MACs(G/s) & \multicolumn{4}{c|}{Kitchen} & \multicolumn{4}{c}{Mall}\\
    \cmidrule(lr){3-6} \cmidrule(lr){7-10}
     &  & t1 & t2 & t3  & avg & t1 & t2 & t3  & avg\\
    \midrule
    1. Noisy  & —— & 6.15 & 6.62 & 6.12 & 6.30  & 45.63 & 58.20 & 62.05 & 55.29  \\
    \midrule
    2. BSRNN  & 1.84 & 5.84 & 6.15 & 5.78 & 5.92 & 39.11 & 50.07 & 55.22 & 48.13 \\
    \midrule
    3. +GR  & 1.09 & 5.92 & 6.24 & 5.90 & 6.02 & 40.11 & 53.10 & 58.04 & 50.42 \\
    \midrule
    4a. + LWR-ASYNC(4) & 1.19 & 5.85 & 6.16 & 5.79 & 5.93 & 39.38 & 50.26 & 55.70 & 48.45   \\
    4b. + LWR-ASYNC(16) & 1.03 & 5.87 & 6.18 & 5.79 & 5.95 & 39.58 & 51.15 & 56.86 & 49.20   \\
    \midrule
    5. ++SBP-P & 0.99 & 5.88 & 6.19 & 5.83 & 5.97 & 39.74 & 52.08 & 56.90 & 49.57  \\ 
    \midrule
    6. +++GR & 0.62 & 5.92 & 6.26 & 5.92 & 6.03 & 40.12 & 53.44 & 58.06 & 50.54   \\ 
    \specialrule{1.2pt}{0pt}{0pt}
    \end{tabular}
\end{adjustbox}
    \label{tab:honor}
    \vspace{-2mm}
\end{table}

Comparing the results in rows 2 and 5 of Table \ref{tab:honor}, we observe that combining the proposed frame resampling with sub-band pruning results in a significant reduction in computational cost, while maintaining ASR performance. In the final row, we demonstrate that our proposed method is compatible with grouped RNN, further reducing computational cost without any degradation in ASR performance, which is consistent with results in Table \ref{tab:dns} and \ref{tab:chime}.